\documentclass[11pt,letterpaper]{JHEP3}
\usepackage{graphics}
\usepackage{epsfig}
\usepackage{amssymb}
\usepackage{stmaryrd}
\usepackage{amsmath}
\usepackage{amsfonts}
\usepackage{mathrsfs}
\usepackage{graphicx}

\def\be{\begin{equation}}
\def\ee{\end{equation}}
\def\ba{\begin{eqnarray}}
\def\ea{\end{eqnarray}}

\title{Periodically Driven Holographic Superconductor}

\author{Wei-Jia Li\\
Center for Theoretical Physics, Massachusetts Institute of Technology,\\
Cambridge, MA 02139, USA\\
Department of Physics, Beijing Normal University, Beijing 100875, China\\
\email{weijiali@mit.edu}}

\author{Yu Tian\\
School of Physics, University of Chinese Academy of Sciences, Beijing 100049, China\\
State Key Laboratory of Theoretical Physics, Institute of Theoretical Physics, \\
Chinese Academy of Sciences, Beijing 100190, China\\
 \email{ytian@ucas.ac.cn}}
 
\author{Hongbao Zhang\\
Theoretische Natuurkunde, Vrije Universiteit Brussel and \\
The International Solvay Institutes,\\
Pleinlaan 2, B-1050 Brussels, Belgium\\
\email{hzhang@vub.ac.be}}

\abstract{As a first step towards our holographic investigation of the far-from-equilibrium physics of periodically driven systems at strong coupling, we explore the real time dynamics of holographic superconductor driven by a monochromatically alternating electric field with various frequencies. As a result, our holographic superconductor is driven to the final oscillating state, where the condensate is suppressed and the oscillation frequency is controlled by twice of the driving frequency. In particular, in the large frequency limit, the three distinct channels towards the final steady state are found, namely under damped to superconducting phase, over damped to superconducting and normal phase, which can be captured essentially by the low lying spectrum of quasi-normal modes in the time averaged approximation, reminiscent of the effective field theory perspective. 
}

\begin{document}
\section{Introduction}
Recent experimental breakthroughs in atomic physics, quantum optics and nano-science have made it possible to manipulate the system at 
hand out of equilibrium in a controllable way. This has been spurring a lot of theoretical investigations of non-equilibrium 
physics in condensed matter systems. On the other hand, the study of non-equilibrium behavior is also highly significant in other fields 
of physics like high energy physics and the physics of the early universe, where the stimulus is being provided in particular by the 
ongoing LHC and Planck experiments respectively. All of these are making the study of non-equilibrium physics a very exciting field 
with various rapid developments. 
 
It is noteworthy that some systems involved in non-equilibrium experiments are strongly coupled. Two prestigious examples are quark 
gluon plasma produced in RHIC as well as LHC experiments and cold atom gases trapped in optical lattices. A suitable tool for 
describing the dynamics of strongly coupled systems is the string inspired AdS/CFT correspondence, where as kind of strong weak 
duality the problem of a strongly coupled system can be holographically mapped to a much simpler calculation in the 
corresponding bulk gravity with one extra dimension. To be more specific, the static black hole in the bulk corresponds to the boundary 
system in equilibrium with finite temperature, where the small perturbation around the black hole induces the boundary system in 
near-equilibrium state such that the linear approximation is reliable on both sides as it should be the case. Over the last few 
years, such a picture has offered us remarkable insights in our understanding various universal equilibrium and near-equilibrium 
properties of strongly coupled systems. On the other hand, in order to address the universal far-from-equilibrium behaviors of strongly coupled systems, 
one is required by holography to have a non-trivial time dependent solution to the non-linear bulk differential equations in question. 

Stimulated by this, a wealth of effort has been made to model non-equilibrium physics by holography\cite{CY,MKT,DNT,AJ,EH,Balasubramanian1,Balasubramanian2,HJW1,GPZ,BD,BBCCF,KKVT,HPG,HJW2,GS,CK,BLM,BGSSW,BDDN,ACL,GGGZZ,BGS,CKY,Balasubramanian3,BLMN,NNT,CKPT}. Among them, most of work focuses on the situation in which the system is driven out of equilibrium by sudden perturbations, namely quantum quenches. However, as we know, there are other various imaginable non-equilibrium protocols. One such possibility is provided by external forces that act on the given system periodically. To the best of our knowledge, the holographic investigation of such a periodic driving has been touched upon only for the holographic superconductor in \cite{BDST}, where it is claimed that the transition temperature is enhanced when the driving frequency is large enough for a sinusoidally varying chemical potential. But nevertheless what is concerned there is essentially the final steady state in the large frequency limit, by which the bulk dynamics can be reduced to an effective static system with the time averaged approximation, while neither the real time dynamics for how the system evolves towards the final steady state nor the regime for small driving frequencies are addressed. The purpose of the present paper is to make a first step towards addressing the real time dynamics of periodically driven strongly coupled systems by holography. To be more precise, we shall see what happens to the holographic superconductors driven by a monochromatically alternating electric field, which is supposed to be more realistic than the varying chemical potential considered in \cite{BDST}.

The rest of our paper is structured as follows. In the subsequent section, we provide a brief review of holographic superconductor in the infalling Eddington coordinates as well as the holographic setup for our periodically driven holographic superconductor by an alternating electric field. Then we present our numerical result for the real time dynamics of condensate in Section \ref{numerics}, where we also show that in the large frequency limit the late time behavior can be well captured by the lowest lying quasi-normal mode in the time averaged approximation. We conclude with some discussions in the final section.
\section{Holographic setup}
To study a two dimensional charged holographic superconductor with a global $U(1)$ symmetry, let us start with the following bulk action of gravity coupled to a $U(1)$ gauge field $A$ and a scalar field $\Psi$ with charge $q$ and mass $m$\cite{HHH1,HHH2}, i.e.,
\begin{equation}\label{bulkaction}
S=
\int_\mathcal{M} d^4x\sqrt{-g}[R+\frac{3}{L^2}+\frac{1}{q^2}(-\frac{1}{4}F_{ab}F^{ab}-|D\Psi|^2-m^2|\Psi|^2)],
\end{equation}
where
 $L$ is the AdS radius, and $D=\nabla-iA$ with $\nabla$ the metric covariant derivative operator. In what follows, we shall work in the probe limit in the sense that the back reaction of matter fields onto the metric can be neglected. Obviously, such a probe limit can be achieved by taking the large $q$ limit. With such a limit, we can put the matter fields onto the vacuum AdS black hole background, which can be written in the infalling Eddington coordinates as
\begin{equation}
ds^2=\frac{L^2}{z^2}[-f(z)dt^2-2dtdz+dx^2+dy^2],
\end{equation}
 where the blackening factor is given by $f(z)=1-(\frac{z}{z_h})^3$ with $z=z_h$ the position of horizon and $z=0$ the AdS boundary. The behavior of matter fields is governed by the following equations of motion, i.e., 
 \begin{equation}
 D_aD^a\Psi-m^2\Psi=0, \nabla_aF^{ab}=i[\Psi^* D^b\Psi-\Psi (D^b\Psi)^*].
 \end{equation}
By the holographic dictionary, the temperature of dual boundary system is given by Hawking temperature as
\begin{equation}
T=\frac{3}{4\pi z_h}.
\end{equation}
The bulk  gauge field $A$ evaluated at the boundary serves as the source for a conserved current $J$ associated with a global $U(1)$ symmetry, and the near boundary data of the scalar field $\Psi$ sources a scalar operator $O$ with the conformal scaling dimension $\Delta=\frac{3}{2}\pm\sqrt{\frac{9}{4}+m^2L^2}$. For simplicity but without loss of generality, we shall be concerned only with the particular case of $m^2L^2 = -2$, which yields an operator of dimension two in the standard quantization. To be more precise, the asymptotic solution of $A$ and $\Psi$ can be expanded near the AdS boundary as 
\begin{eqnarray}
&&A_\nu=a_\nu+b_\nu z+o(z),\\
&&\Psi=\frac{1}{L}[\phi z+z^2\varphi+o(z^2)],
\end{eqnarray}
where we have been working in the axial gauge $A_z=0$.
 Then by the standard AdS/CFT prescription, the expectation value of the corresponding boundary quantum field theory operators $\langle J\rangle$ and $\langle O\rangle$ can be obtained by variation of the action with respect to the sources, i.e., 
 \begin{eqnarray}
&& \langle J^\nu\rangle=\frac{\delta S_{ren}}{\delta a_\nu}=\lim_{z\rightarrow 0}\frac{\sqrt{-g}}{q^2}F^{z\nu},\\
&&\langle O\rangle=\frac{\delta S_{ren}}{\delta\phi}=\lim_{z\rightarrow 0}[\frac{z\sqrt{-g}}{Lq^2}(D^z\Psi)^*-\frac{z\sqrt{-\gamma}}{L^2q^2}\Psi^*]=\frac{1}{q^2}(\varphi^*-\dot{\phi^*}-ia_t\phi^*),
 \end{eqnarray}
 where $S_{ren}$ is the renormalized action obtained by adding the counter term to the original action, i.e.,
 \begin{equation}
 S_{ren}=S-\frac{1}{Lq^2}\int_\mathcal{B}\sqrt{-\gamma}|\Psi|^2,
 \end{equation}
 and the dot denotes the derivative with respect to the time $t$. 
 
\begin{figure}
\center{
\includegraphics[scale=0.7]{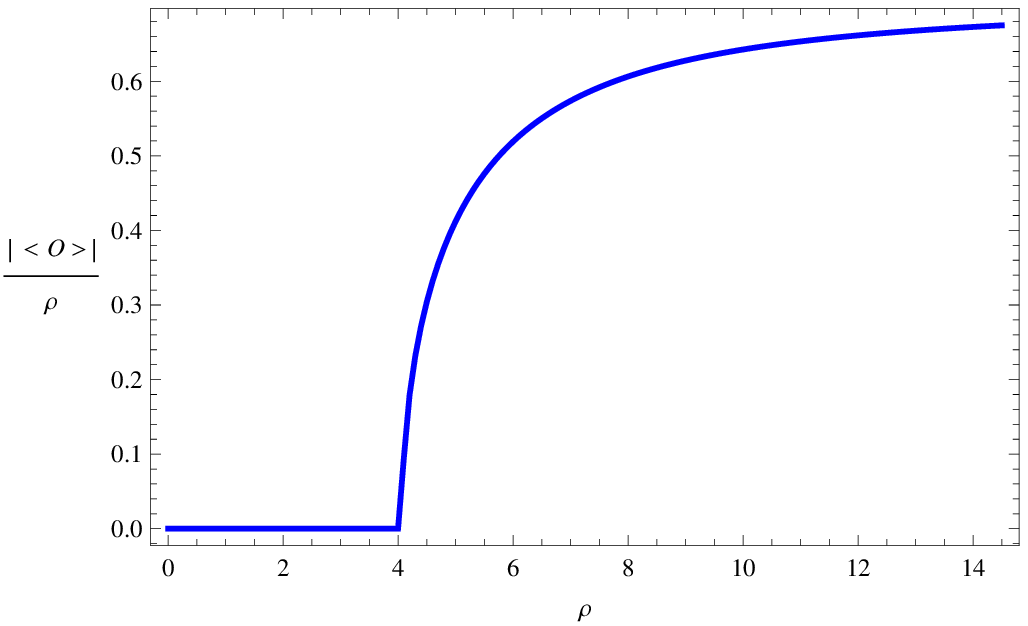}\hspace{1cm}
\includegraphics[scale=0.8]{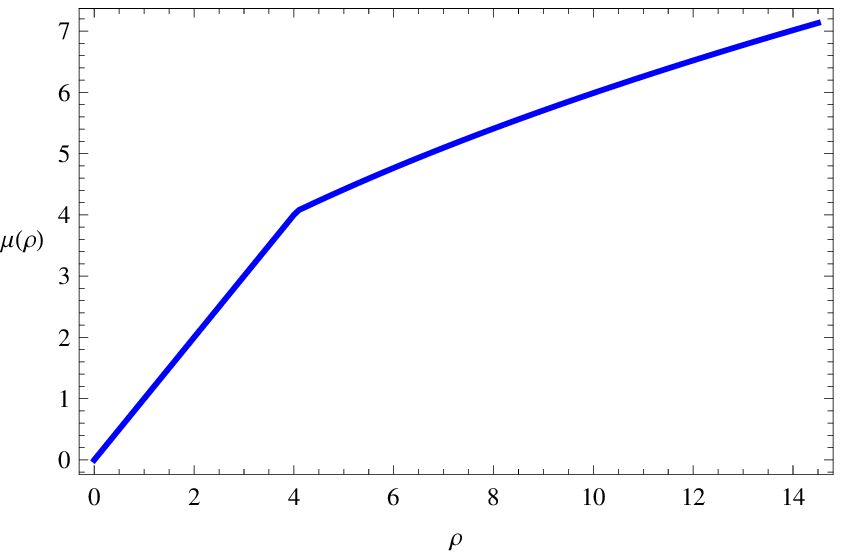}\hspace{1cm}
\caption{\label{condensate} The condensate and chemical potential as a function of  charge density with the critical charge density $\rho_c=4.064$.}}
\end{figure}

\begin{figure}
\center{
\includegraphics[scale=0.8]{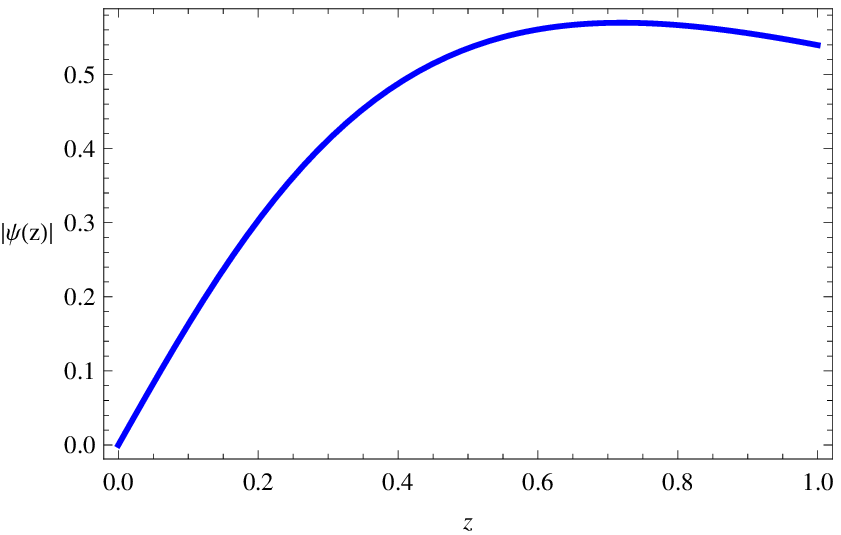}\hspace{1cm}
\includegraphics[scale=0.8]{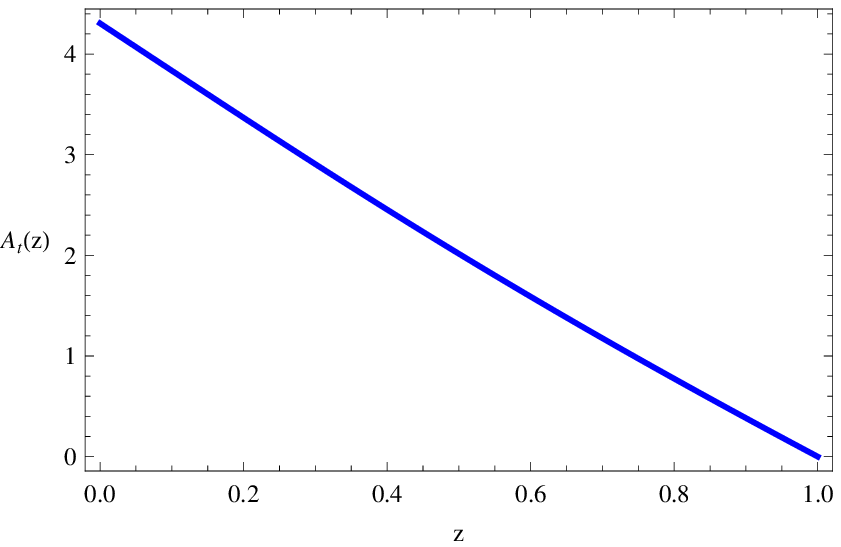}\hspace{1cm}
\caption{\label{profile} The profile of amplitude of scalar field and electromagnetic potential for the superconducting phase at the charge density $\rho=4.7$.}}
\end{figure}

From here on we shall work in units in which $L=1$, $q=1$, and $z_h=1$. To proceed, we would like to write down the equations of motion in an explicit way as
\begin{eqnarray}
0&=&2\partial_t\partial_z\psi+\frac{2}{z^2}f\psi-\frac{f'}{z}\psi-f'\partial_z\psi-f\partial_z^2\psi-i\partial_zA_t\psi-2iA_t\partial_z\psi\nonumber\\
&&-\partial_x^2\psi-\partial_y^2\psi+i\partial_xA_x\psi+i\partial_yA_y\psi+2iA_x\partial_x\psi+2iA_y\partial_y\psi\nonumber\\
&&+A_x^2\psi+A_y^2\psi-\frac{2}{z^2}\psi
\end{eqnarray}
for the Klein-Gordon equation with $\Psi=z\psi$ and
\begin{equation}
\partial_z^2A_t-\partial_z\partial_xA_x-\partial_z\partial_yA_y=i(\psi^*\partial_z\psi-\psi\partial_z\psi^*),
\end{equation}
\begin{eqnarray}
&&\partial_t\partial_zA_t+\partial_t\partial_xA_x+\partial_t\partial_yA_y-\partial_x^2A_t-\partial_y^2A_t-f\partial_z\partial_xA_x-f\partial_z\partial_yA_y\nonumber\\
&&=-i(\psi^*\partial_t\psi-\psi\partial_t\psi^*)-2A_t\psi^*\psi+if(\psi^*\partial_z\psi-\psi\partial_z\psi^*),
\end{eqnarray}
\begin{eqnarray}
&&\partial_z\partial_xA_t+f\partial_z^2A_x+f'\partial_zA_x-2\partial_t\partial_zA_x+\partial_y^2A_x-\partial_x\partial_yA_y\nonumber\\
&&=i(\psi^*\partial_x\psi-\psi\partial_x\psi^*)+2A_x\psi^*\psi \  \ (x\leftrightarrow y)
\end{eqnarray}
for the Maxwell equations with the first one as the constraint equation, where the prime denotes the differentiation with respect to $z$.
To make our life easier, from now on we shall make the following ansatz, i.e.,
\begin{equation}
A_y=\partial_xA_t=\partial_yA_t=\partial_xA_x=\partial_yA_x=\partial_x\psi=\partial_y\psi=0.
\end{equation}
With this, the equations of motion are simplified dramatically as
\begin{eqnarray}
&&2\partial_t\partial_z\psi+\frac{2}{z^2}f\psi-\frac{f'}{z}\psi-f'\partial_z\psi-f\partial_z^2\psi-i\partial_zA_t\psi-2iA_t\partial_z\psi+A_x^2\psi-\frac{2}{z^2}\psi=0,\label{KG}\\
&&\partial_z^2A_t=i(\psi^*\partial_z\psi-\psi\partial_z\psi^*),\\
&&\partial_t\partial_zA_t=-i(\psi^*\partial_t\psi-\psi\partial_t\psi^*)-2A_t\psi^*\psi+if(\psi^*\partial_z\psi-\psi\partial_z\psi^*),\\
&&f\partial_z^2A_x+f'\partial_zA_x-2\partial_t\partial_zA_x=2A_x\psi^*\psi.
\end{eqnarray}
As we know, the equilibrium state on the boundary, by holography, corresponds to the following static solution in the bulk, i.e.,
\begin{equation}
A_x=\partial_t\psi=\partial_tA_t=0,
\end{equation} 
whereby it is not hard to see there exist a trivial solution, i.e.,
 \begin{equation}
\psi=0, A_t=\mu(1-z).
 \end{equation}
This solution corresponds to the normal phase on the boundary with $\mu$ the chemical potential as well as the charge density.  On the other hand, it is generically impossible to find the corresponding analytic static bulk solution for the superconducting phase on the boundary, so here comes the numerics. First note that the above equations of motion imply the following boundary condition
 \begin{equation}
 A_t=0, 2\psi\psi^*+3(\psi\psi^*)'=0
 \end{equation}
 on the horizon. Then in order to see whether the dual boundary system is spontaneously broken to the superconducting phase at the given charge density, a natural boundary condition at the AdS boundary is imposed as follows
 \begin{equation}
 A_t'=-\rho,\psi=0.
 \end{equation}
 Now with the above two sets of boundary conditions accompanied by the gauge fixing $\psi=\psi^*$ at the horizon, we can resort to the pseudo-spectral method to solve the equations of motion.
The relevant results are plotted in Fig.\ref{condensate} and Fig.\ref{profile}. As one can see, the condensate is spontaneously formed when one increases the charge density above the critical value $\rho_c=4.064$, which corresponds to a nontrivial static solution in the bulk.

Next let us see what happens to the above holographic superconductor at the fixed charged density when we irradiate it with an alternating electric field as
\begin{equation}
E_x=E\cos(\omega t), t\geq0,
\end{equation}
which can be achieved by the following boundary condition
\begin{equation}
A_t'=-\rho, A_x=-\frac{E}{\omega}\sin(\omega t), t\geq 0
\end{equation}
at the AdS boundary.
With the initial configuration given by the equilibrium state at $t=0$ and subject to the aforementioned boundary condition at the AdS boundary, the equations of motion can be solved numerically in the gauge $A_t=0$ on the horizon by resorting to the pseudo-spectral method along the radial $z$ direction and the forth order Runge-Kutta method in the time $t$. The relevant numerical results will be presented in the next section.
\section{Numerical results\label{numerics}}
\subsection{Real time dynamics towards the final state}
\begin{figure}
\center{
\includegraphics[scale=0.7]{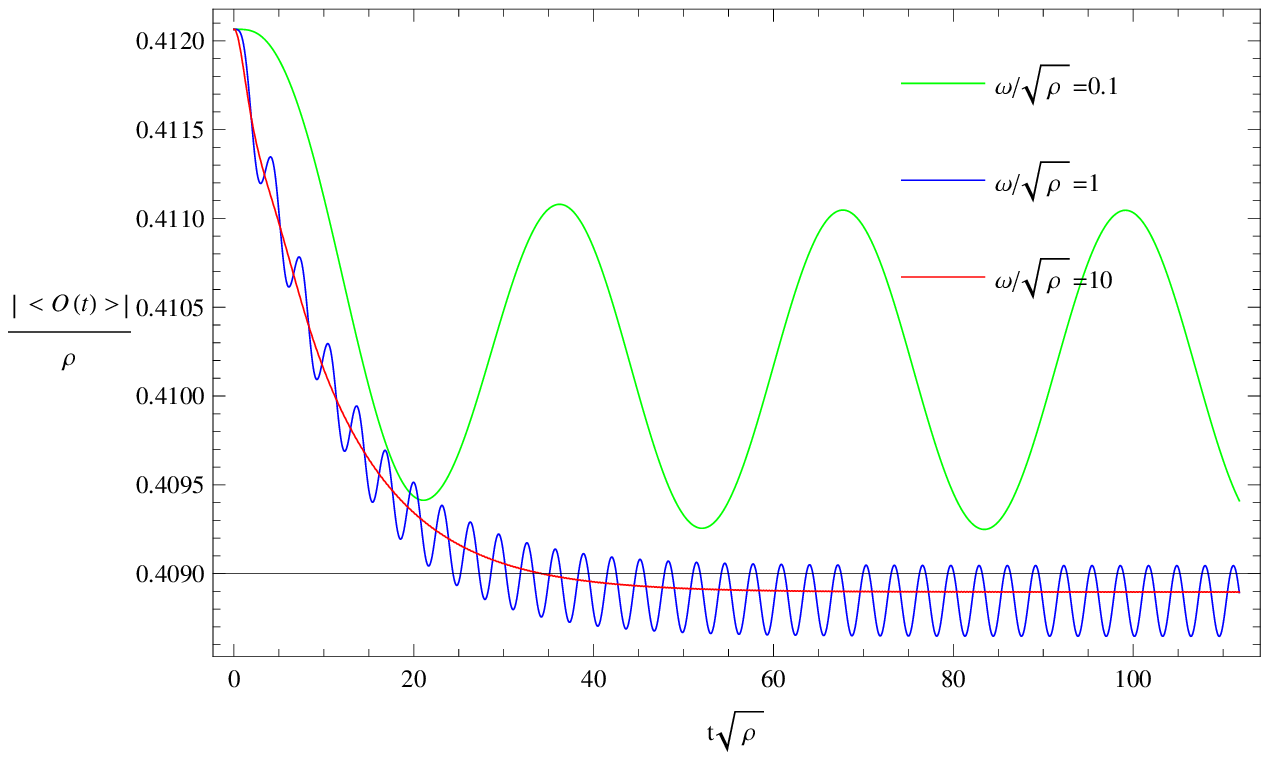}\hspace{1cm}
\includegraphics[scale=0.7]{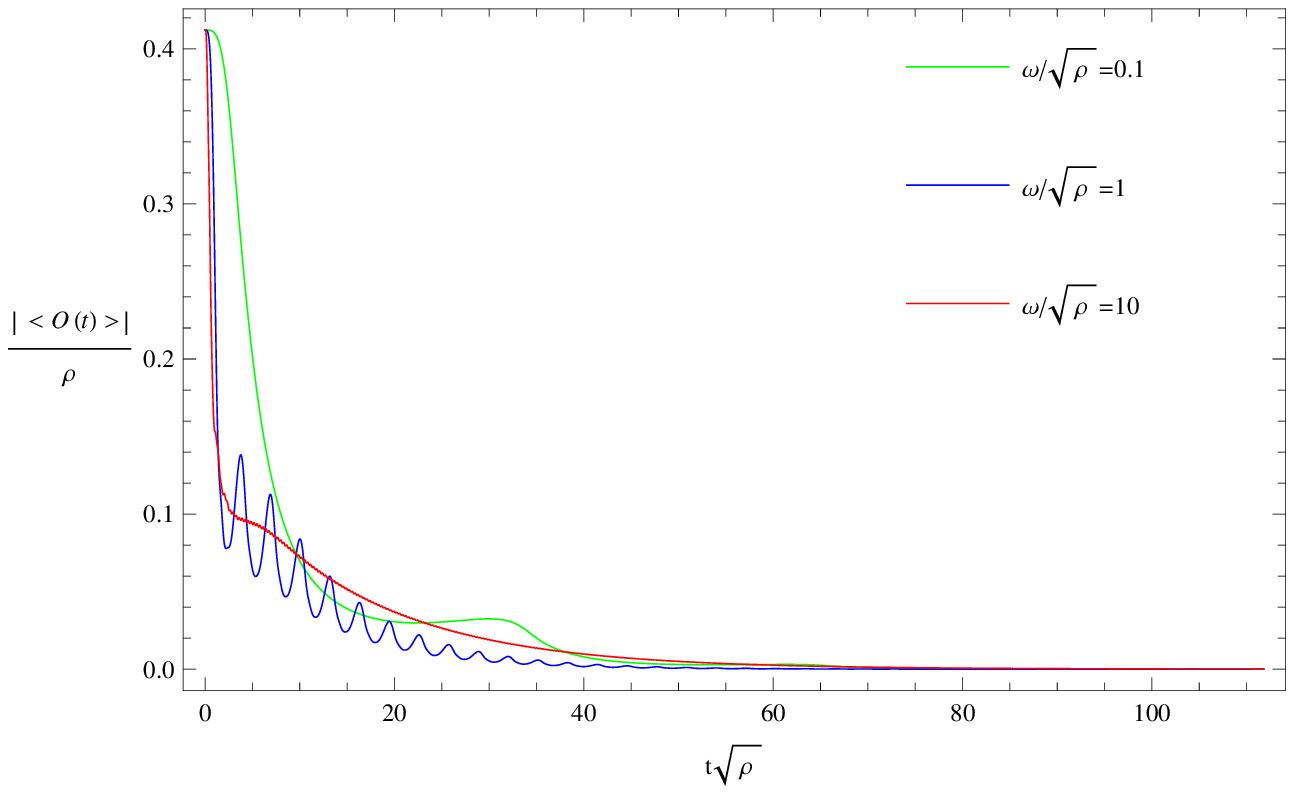}\hspace{1cm}
\caption{\label{realtime5} The real time dynamics of condensate for the charge density $\rho=5$, where the upper panel is for $\frac{E}{\omega\sqrt{\rho}}=0.1$ and the lower panel is for $\frac{E}{\omega\sqrt{\rho}}=5$.}}
\end{figure}

\begin{figure}
\center{
\includegraphics[scale=0.65]{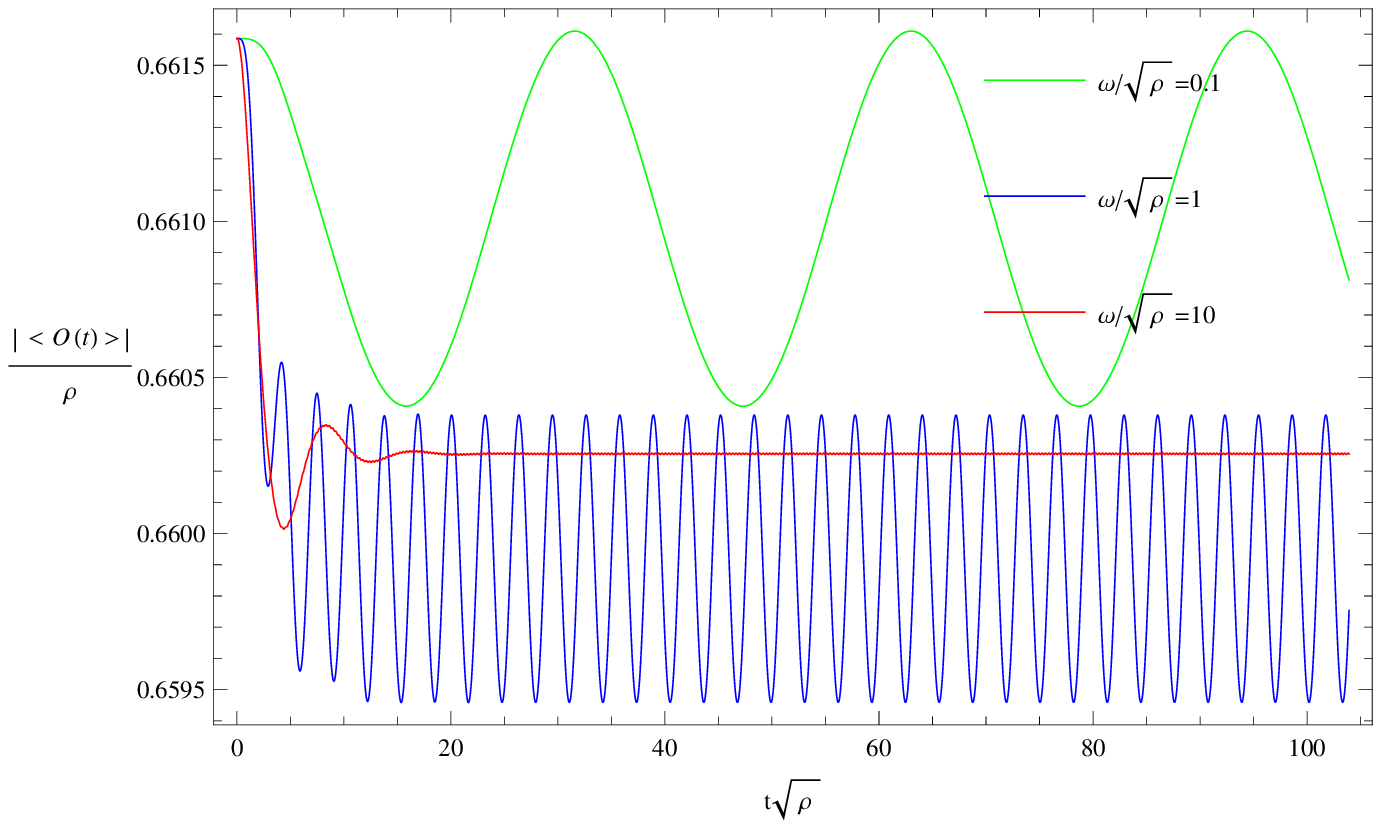}\hspace{1cm}
\includegraphics[scale=0.6]{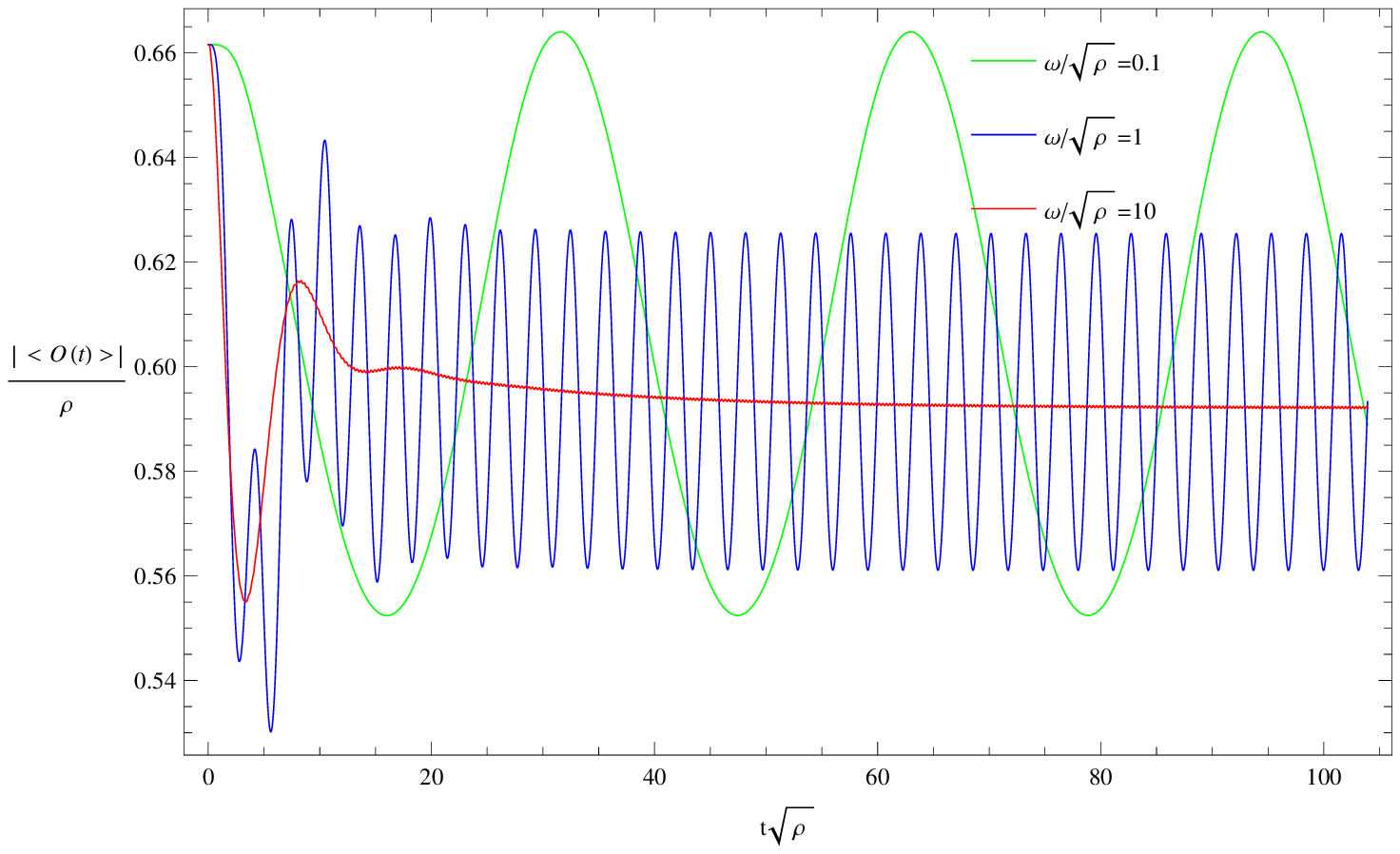}\hspace{1cm}
\caption{\label{realtime12} The real time dynamics of condensate for the charge density $\rho=12$, where the upper panel is for $\frac{E}{\omega\sqrt{\rho}}=0.1$ and the lower panel is for $\frac{E}{\omega\sqrt{\rho}}=1$.}}
\end{figure}

Let us start to demonstrate our numerical results with Fig.\ref{realtime5} and Fig.\ref{realtime12} as typical examples of the real time dynamics for our condensate. As we shall see below, similar results are also expected for the other values of charge density we have experimented with.

As illustrated, the condensate is suppressed under the alternating electric field, which is reasonable because such an electric field increases the effective mass of the charged scalar field as shown in the Klein Gordon equation (\ref{KG}). This observation is further consistent with the phenomenon that the condensate is decreased with the increase of the driving amplitude $\frac{E}{\omega}$.
In addition, after a transient response, the condensate ends up with an oscillating state. The resulting oscillation frequency turns out to be twice of the driving frequency, which comes essentially from the fact that the Klein Gordon equation (\ref{KG}) is quadratic in $A_x$. 

\begin{figure}
\center{
\includegraphics[scale=1.2]{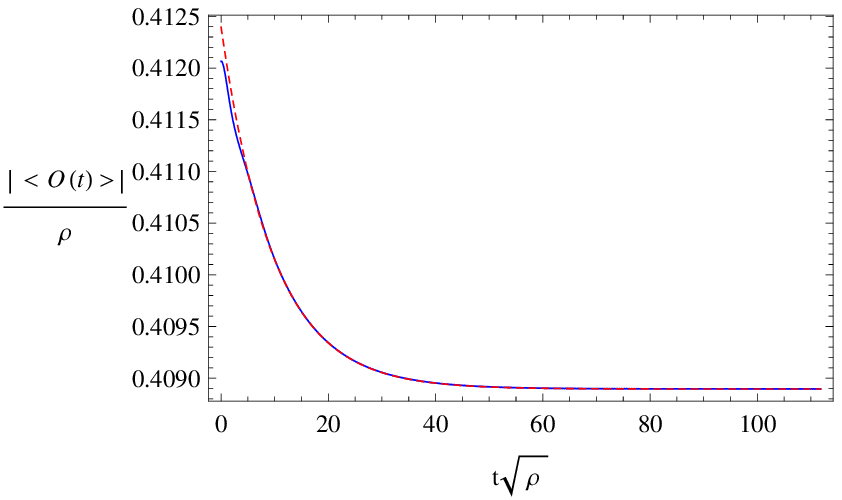}\hspace{1cm}
\includegraphics[scale=1.2]{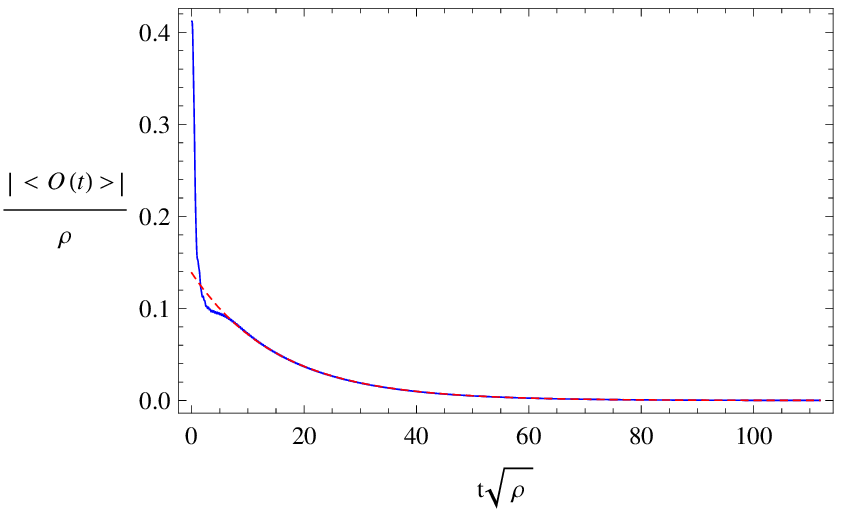}\hspace{1cm}
\caption{\label{fitting5} The late time behavior can be well fit by an exponential decay for the charge density $\rho=5$ and the driving frequency $\frac{\omega}{\sqrt{\rho}}=10$, where the decay rate is given as $0.103$ for $\frac{E}{\omega\sqrt{\rho}}=0.1$  in the upper panel and $0.066$ for $\frac{E}{\omega\sqrt{\rho}}=5$ in the lower panel.}}
\end{figure}

\begin{figure}
\center{
\includegraphics[scale=0.95]{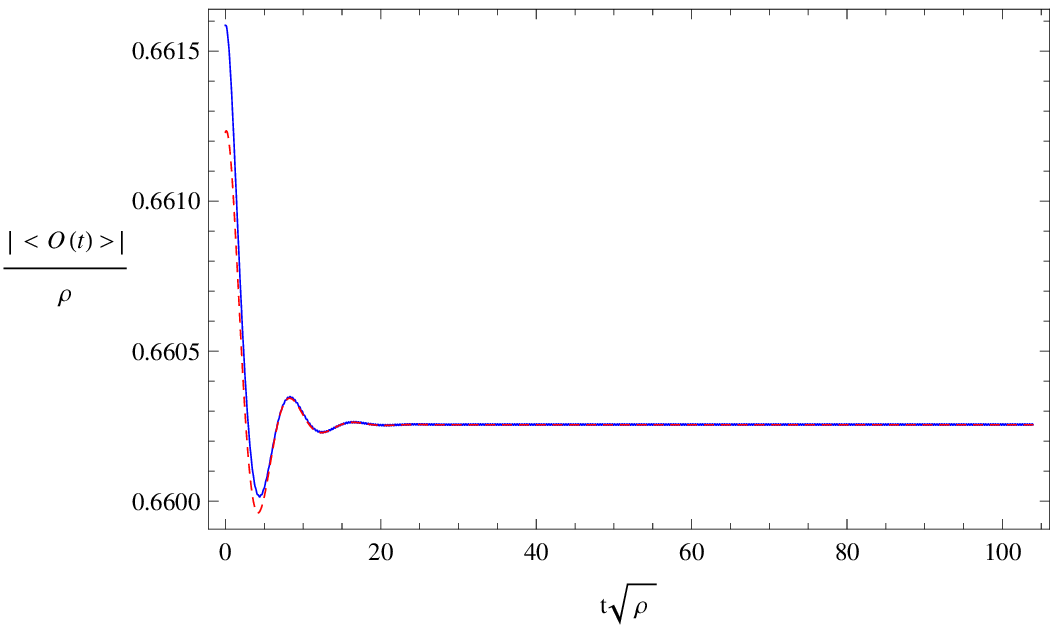}\hspace{1cm}
\includegraphics[scale=1.0]{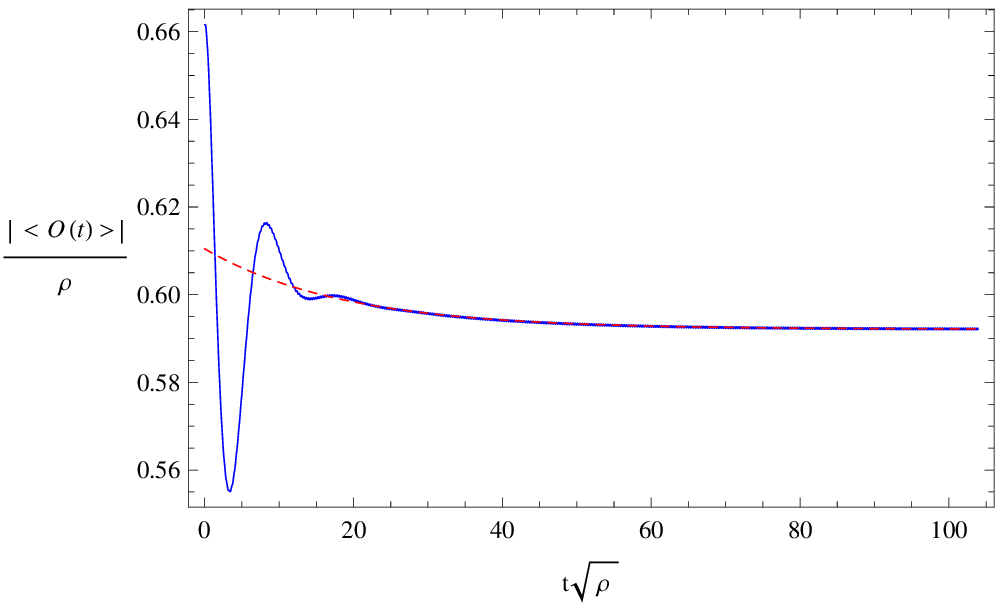}\hspace{1cm}
\caption{\label{fitting12} The fitting plot of the late time behavior for the charge density $\rho=12$ and the driving frequency $\frac{\omega}{\sqrt{\rho}}=10$.  For the case of $\frac{E}{\omega\sqrt{\rho}}=0.1$, as shown in the upper panel, the late time behavior can be well fit by an under damped mode with the decay rate $0.295$ and the oscillation frequency 0.770.  For the case of $\frac{E}{\omega\sqrt{\rho}}=1$, as shown in the lower panel, the late time behavior can be well fit by an over damped mode with the decay rate given by $0.054$.}}
\end{figure}

On the other hand, the oscillation amplitude is decreased with the increase of the driving frequency. In particular, the oscillation amplitude goes to zero in the large frequency limit, which amounts to saying that the corresponding condensate approaches to a final steady state. Furthermore, as shown in Fig.\ref{fitting5} and Fig.\ref{fitting12}, the late time behavior can be well captured by
\begin{equation}\label{fittingbroken}
| \langle O(t)\rangle|=|\langle O_f\rangle|+\delta e^{-i\Omega t}+\delta^*e^{i\Omega^* t}
\end{equation}
for the final condenstate $\langle O_f\rangle\neq0$ and
\begin{equation}\label{fittingnormal}
| \langle O(t)\rangle|=|\langle O_f\rangle+\delta e^{-i\Omega t}|=|\delta| e^{\mathrm{Im}(\Omega)t}
\end{equation}
for $\langle O_f\rangle=0$ if appropriate $\Omega$ and $\delta$ are chosen.
For the holographic superconductor with a small charge density, the late time behavior can be well fit by an exponential decay, where our holographic superconductor will be over damped to the normal state when the driving amplitude is larger than a critical value $(\frac{E}{\omega})_c$. Compared with this, when our holographic superconductor is at a fixed charge density larger than a certain value denoted by $\rho_*$ and driven by a small driving amplitude, the late time evolution displays an under damped behavior, which turns to an over damped one if the driving amplitude is increased beyond a certain value $(\frac{E}{\omega})_*$.

In summary,  as a response to the alternating electric field, our holographic superconductor approaches generically to the final oscillating state with the condensate suppressed and the oscillation frequency twice of the driving frequency. In particular, in the large frequency limit, the late time dynamics demonstrates the three distinct routes towards the final steady state, namely under damped to superconducting phase, over damped to superconducting and normal phase. Such an intriguing late time behavior begs for a deeper understanding. As we shall show right now in the subsequent subsection, this can actually be well understood by the low lying spectrum of quasi-normal modes in the time averaged approximation, reminiscent of the effective field theory perspective, where the low energy physics can be obtained by integrating out the high energy modes, which effect the low energy behavior only through some finite relevant parameters.
\subsection{Quasi-normal modes in the time averaged approximation}

\begin{figure}
\center{
\includegraphics[scale=0.65]{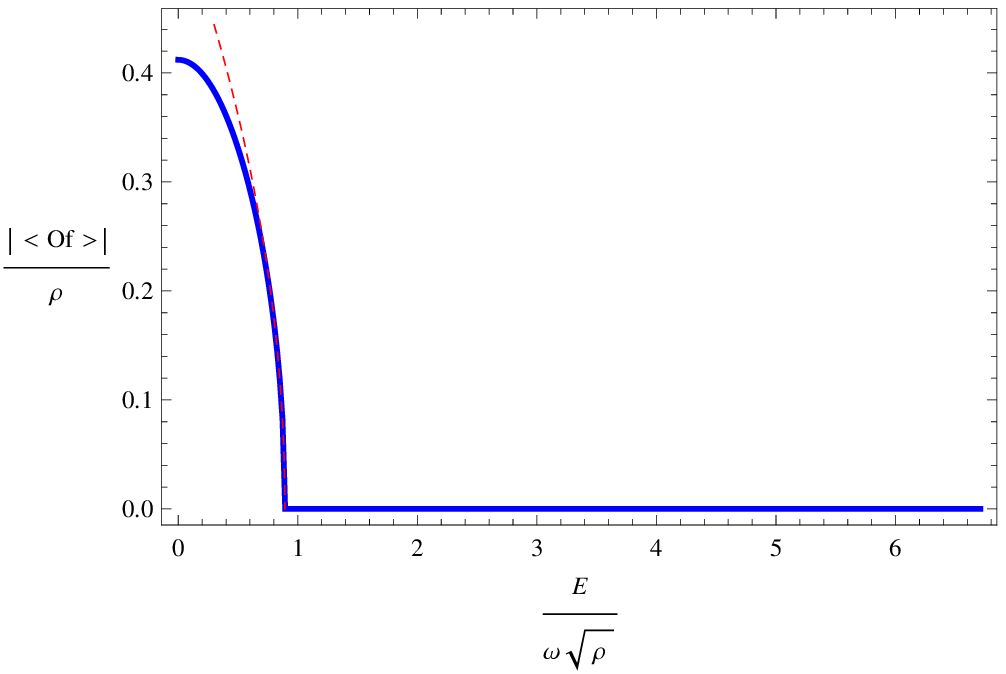}\hspace{1cm}
\includegraphics[scale=0.8]{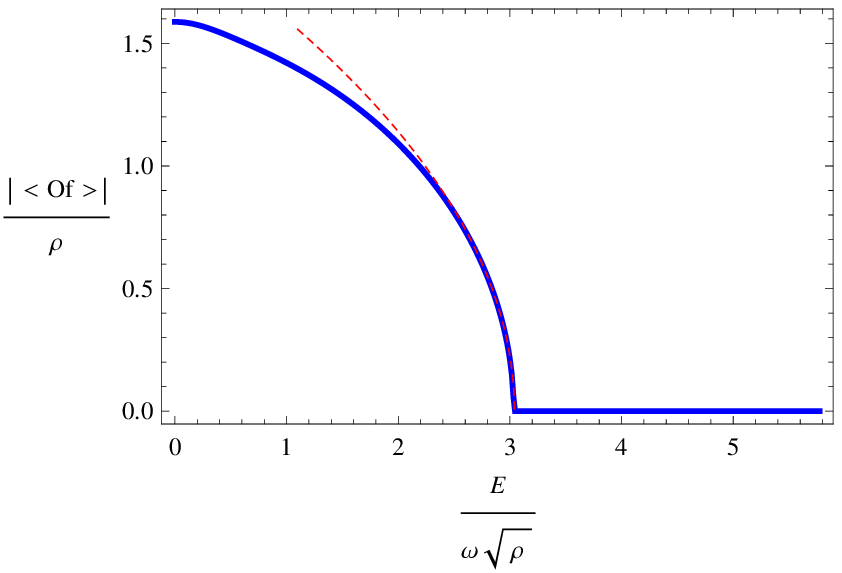}\hspace{1cm}
\caption{\label{newcond} The final condensate as a function of driving amplitude with the charge density fixed, where the left panel is for $\rho=5$ with $(\frac{E}{\omega})_c=0.892\sqrt{\rho}$ and the right panel is for $\rho=12$ with $(\frac{E}{\omega})_c=3.040\sqrt{\rho}$. In both cases, the condensate near the critical point can be well fit by the power law behavior as $|\langle O_f\rangle|\propto[(\frac{E}{\omega})_c-\frac{E}{\omega}]^\frac{1}{2}$ with the proportional coefficient given by $1.930$ for $\rho=5$ and $7.208$ for $\rho=12$.}}
\end{figure}

In order to study the low lying spectrum of quasi-normal modes in the time averaged approximation, we first notice that  the final periodically varying state can actually be reconstructed by Fourier series in the time. In particular, in the large frequency limit, the final state can be captured by the following surviving modes
\begin{equation}
\psi=\psi^f(z),A_t=A_t^f(z),A_x=-\frac{E}{\omega}\sin(\omega t)
\end{equation}
together with the corresponding time averaged equations of motion
\begin{eqnarray}
&&\frac{2}{z^2}f\psi^f-\frac{f'}{z}\psi^f-f'\partial_z\psi^f-f\partial_z^2\psi^f-i\partial_zA_t^f\psi^f-2iA_t^f\partial_z\psi^f+\frac{E^2}{2\omega^2}\psi^f-\frac{2}{z^2}\psi^f=0,\\
&&\partial_z^2A_t^f=i(\psi^{f*}\partial_z\psi^f-\psi^f\partial_z\psi^{f*}),\\
&&0=-2A_t^f\psi^{f*}\psi^f+if(\psi^{f*}\partial_z\psi^f-\psi^f\partial_z\psi^{f*}),
\end{eqnarray}
subject to the boundary conditions
\begin{equation}
A_t^{f'}=-\rho,\psi^f=0
\end{equation}
at the AdS boundary and 
\begin{equation}
 A_t^f=0, (2+\frac{E^2}{\omega^2})\psi^f\psi^{f*}+3(\psi^f\psi^{f*})'=0
\end{equation}
on the horizon. Then by our numerics, we are allowed to obtain the condensate as a function of driving amplitude at the given charge density. As expected and shown in Fig.\ref{newcond}, the resultant condensate is decreased with the increase of the driving amplitude, which has a critical value as denoted previously by $(\frac{E}{\omega})_c$ where our holographic superconductor exhibits a phase transition with $\frac{1}{2}$ as  the critical exponent to the normal state. The corresponding bulk solution to such a normal state is given by
\begin{equation}
\psi^f=0, A_t^f=\rho(1-z).
\end{equation}
\begin{figure}
\center{
\includegraphics[scale=0.5]{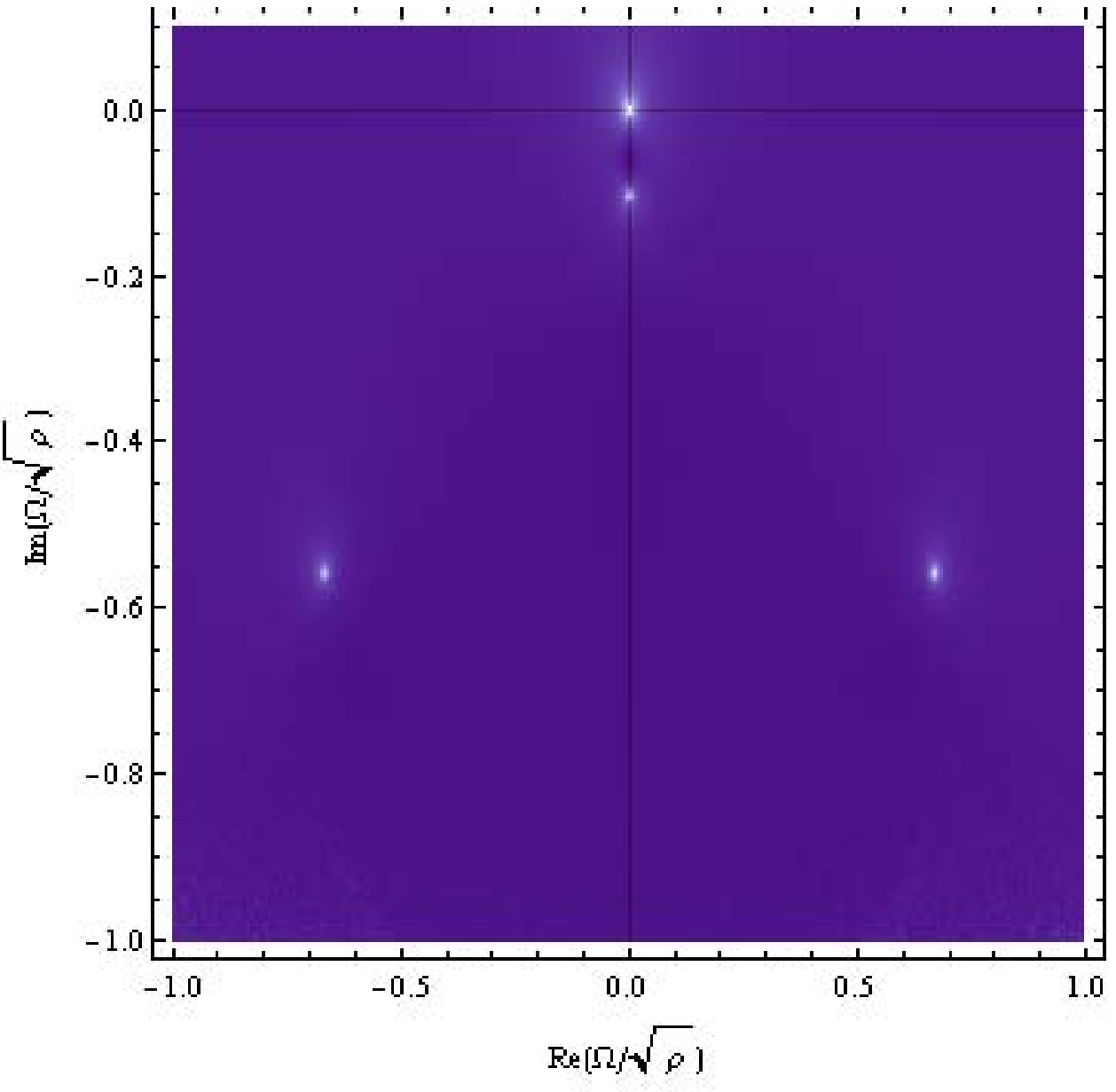}\hspace{1cm}
\includegraphics[scale=0.5]{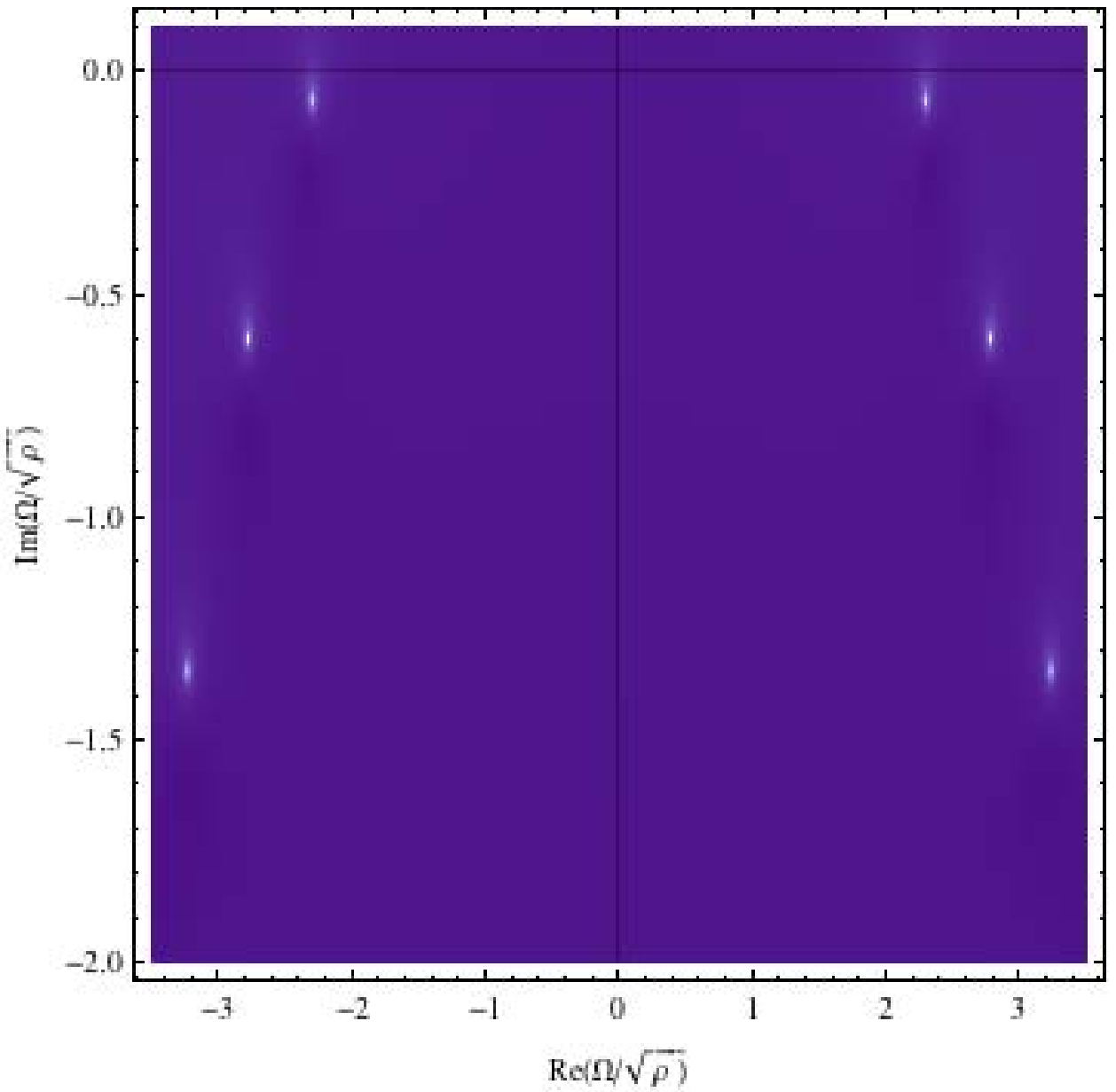}\hspace{1cm}
\caption{\label{qnm5} The low lying quasi-normal modes around the final state captured by the time averaged approximation in the large frequency limit for the charge density $\rho=5$, where the left plot is for $\frac{E}{\omega\sqrt{\rho}}=0.1$ and the right plot is for $\frac{E}{\omega\sqrt{\rho}}=5$.  The plots are generated as density plots of $|\frac{d\mathrm{det}[\mathbf{L}(\Omega)]}{d\Omega}/\mathrm{det}[\mathbf{L}(\Omega)]|$ and the bright spots are identified as locations of quasi-normal modes. For the case of $\frac{E}{\omega\sqrt{\rho}}=0.1$, which ends up with the superconducting phase as the final state, besides the Goldstone mode pinned at the origin, the lowest lying quasi-normal mode is the Higgs mode sitting at the imaginary axis with $\frac{\Omega}{\sqrt{\rho}}=-0.103i$. For the case of $\frac{E}{\omega\sqrt{\rho}}=5$, which has the normal phase as the final state, the Goldstone disappears as it should be the case with the lowest lying quasi-normal modes  $\frac{\Omega}{\sqrt{\rho}}=\pm 2.305-0.066i$. Both results are in good agreement with the late time behavior following from the real time dynamics. }}
\end{figure}

Now let us move onto the low lying quasi-normal modes on top of such a background solution. Regarding the background solution dual to the above normal state, the corresponding linear perturbation equations are decoupled. For our purpose, we focus exclusively with the linear perturbation of the charged scalar field, where the quasi-normal modes are determined by the following gauge invariant linear perturbation equation\cite{AKL}
\begin{equation}
2\partial_t\partial_z\delta\psi+\frac{2}{z^2}f\delta\psi-\frac{f'}{z}\delta\psi-f'\partial_z\delta\psi-f\partial_z^2\delta\psi+i\rho\delta\psi-2i\rho(1-z)\partial_z\delta\psi+\frac{E^2}{2\omega^2}\delta\psi-\frac{2}{z^2}\delta\psi=0,
\end{equation}
if we set $\delta\psi=e^{-i\Omega t}\delta(z)$ with the non-trivial profile $\delta(z)$ regular on the horizon and vanishing at the AdS boundary. Numerically, by the pseudo-spectral method, the above equation can be cast into a linear algebraic equation as $\mathbf{L}(\Omega, \rho, \frac{E}{\omega})\mathbf{v}=0$ with $\mathbf{L}$ a matrix and $\mathbf{v}$
a vector consisting of $\delta(z)$ evaluated at the grid points. To have a non-trivial profile $\delta(z)$, the determinant of $\mathbf{L}$ is forced to be vanishing, namely $\mathrm{det}[\mathbf{L}(\Omega,\rho,\frac{E}{\omega})]=0$, which gives rise to the quasi-normal frequencies\cite{BGSSW,Leaver,FHS}. The same numerical scheme can be used to identify the quasi-normal frequencies around the bulk solution dual to the superconducting phase, But nevertheless in this case the linear perturbation equations are coupled. The corresponding gauge invariant linear perturbation equations can be obtained directly from the gauge invariant formulation of bulk dynamics for our quasi-normal modes, which has been relegated in Appendix.

\begin{figure}
\center{
\includegraphics[scale=0.5]{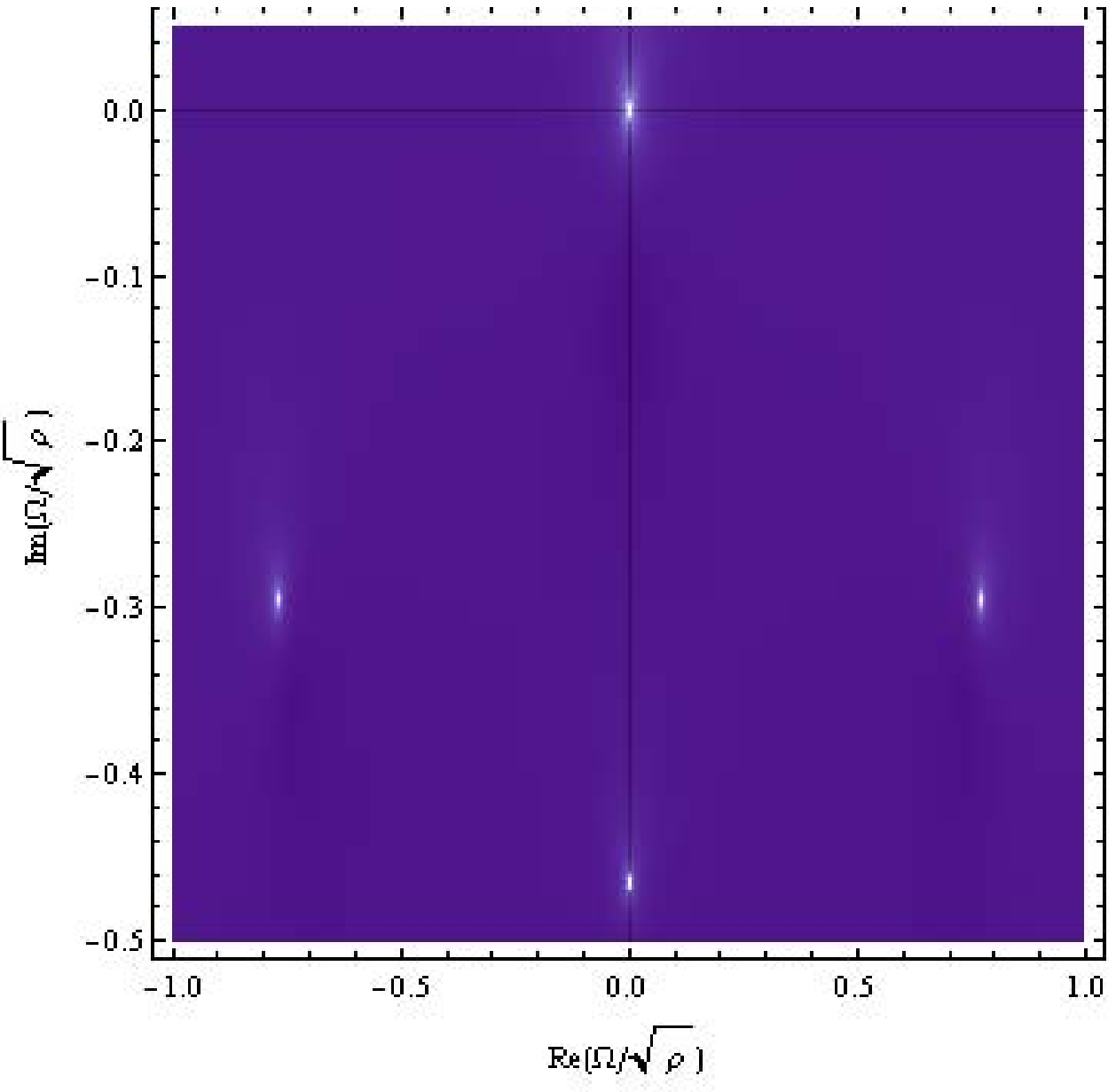}\hspace{1cm}
\includegraphics[scale=0.4]{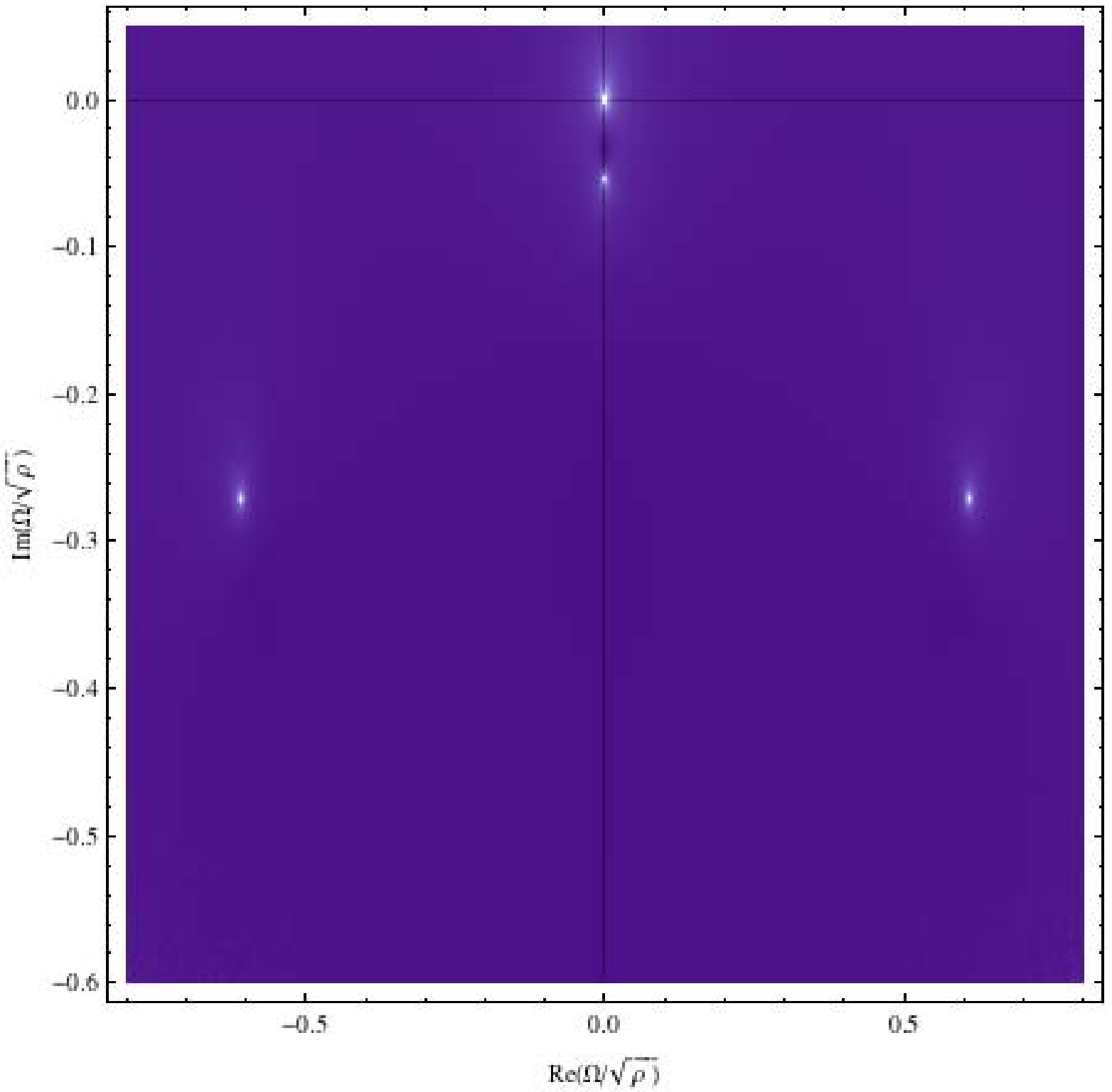}\hspace{1cm}
\caption{\label{qnm12} The low lying quasi-normal modes around the final state captured by the time averaged approximation in the large frequency limit for the charge density $\rho=12$, where the left plot is for $\frac{E}{\omega\sqrt{\rho}}=0.1$ and the right plot is for $\frac{E}{\omega\sqrt{\rho}}=1$.  The Goldstone mode is pinned at the origin for both cases, which corresponds to the superconducting phase as the final state. For the case of $\frac{E}{\omega\sqrt{\rho}}=0.1$,  the lowest lying quasi-normal mode is given by $\frac{\Omega}{\sqrt{\rho}}=\pm 0.770-0.295i$. For the case of $\frac{E}{\omega\sqrt{\rho}}=1$, the lowest lying quasi-normal mode is given by the Higgs mode $\frac{\Omega}{\sqrt{\rho}}=-0.054i$. Both results are in good agreement with the late time behavior following from the real time dynamics. }}
\end{figure}

With the above preparation, we present the relevant numerical results for the corresponding low lying quasi-normal modes in Fig.\ref{qnm5} and Fig.\ref{qnm12}. For the holographic superconductor at the small charged density, the lowest lying mode is the Higgs mode, which is sitting right at the imaginary axis\cite{PAA,PS}. As the alternating electric field is turned on with the driving amplitude increased gradually, this Higgs mode climbs up the imaginary axis and eventually coincides with the Goldstone mode pinned at the origin at the critical driving amplitude $(\frac{E}{\omega})_c$, which corresponds to a phase transition to the normal state. If one cranks up the driving amplitude further, the above two modes begin to depart each other and migrate downwards at the same time, ending up at both sides of the imaginary axis symmetrically. 

On the other hand, when one increases the charge density for our holographic superconductor, the Higgs mode travels down the imaginary axis while the next lowest lying modes, which are located symmetrically at both sides of the imaginary axis, migrate upwards. At some charge density, which is simply $\rho_*$ mentioned in the previous subsection, the next lowest lying modes share the same imaginary part as the Higgs mode. If the charge density is pushed up further, then the Higgs mode steps down from the role of the lowest lying mode, which is now taken over by the originally next lowest lying modes. Now let us turn on the alternating electric field and increase the driving amplitude for our holographic superconductor at such a large charge density. As a result, both the lowest lying mode and Higgs mode migrate upwards. But the Higgs mode is like a rabbit while the lowest lying mode is like a turtle. When the driving amplitude is increased to a certain value, namely $(\frac{E}{\omega})_*$, the Higgs mode catches up with the lowest lying mode. If one cranks up the driving amplitude further, then the Higgs mode wins out and becomes the lowest lying mode, and subsequently the story goes as happens for the case of holographic superconductor at the small charge density.

\begin{figure}
\center{
\includegraphics[scale=0.7]{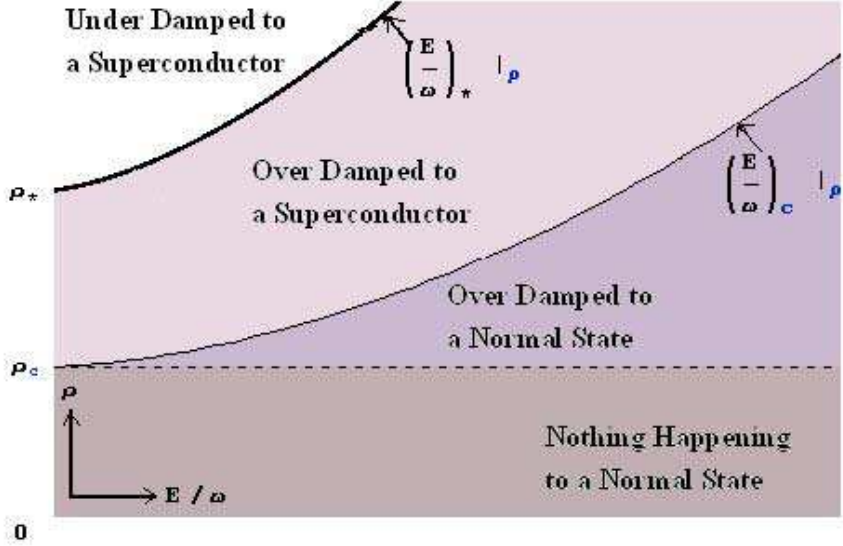}\hspace{1cm}
\caption{\label{diagram} The dynamical phase cartoon diagram towards the final steady state driven periodically by an electric field in the large frequency limit, where $\rho_*$ is around 8.850.}}
\end{figure}

As we have shown in the last subsection, the resultant lowest lying quasi-normal mode described above, substituted into Eq.(\ref{fittingbroken}) and Eq.(\ref{fittingnormal}), provides a remarkably good description of late time behavior for our condensate. Furthermore, by combining the final time averaged bulk background solution with its low lying spectrum of quasi-normal modes, we wind up with the dynamical phase cartoon diagram towards the final steady state, which is depicted in Fig.\ref{diagram}. As we see, the three routes are separated by the two critical lines for our holographic superconductor. One is  $(\frac{E}{\omega})_*|_\rho$ emanating from $\rho_*$. The other is   $(\frac{E}{\omega})_c|_\rho$ emanating from $\rho_c$. As to the normal state, it apparently remains unbroken.

We would like to end this subsection by making a comparison with the previous work \cite{BGSSW}, where the similar late time dynamical phase diagram is obtained by taking into account the back reaction of a homogeneous and isotropic Gaussian-type quench from the scalar field. But due to such a back reaction,  the final black hole temperature turns out to be a monotonically increasing function of the quench strength, which gives rise to an one dimensional dynamical phase diagram rather than the two dimensional one we have obtained for our periodically driven holographic superconductor.

\section{Conclusion}
As promised, we have explored the real time dynamics of our holographic superconductor driven by an alternating electric field. As a result, the holographic superconductor is driven generically to a final oscillating state, where the condensate is suppressed with the oscillation frequency twice of the driving one. In particular, when the driving frequency is large enough, the final oscillating state approaches to a steady state. Furthermore, it is found that there are three distinct late time dynamical behaviors towards the final steady state, namely under damped to superconducting phase, over damped to superconducting and normal phase. Remarkably, such a dynamical phase diagram can be well understood by analyzing the final time averaged bulk background solution as well as its low lying spectrum of quasi-normal modes. Although the relevant results such as the dynamical phase diagram may not be applicable to other periodically driven systems at strong coupling, we believe that the whole strategy we have developed here for our holographic superconductor can also apply to them. 

We conclude with various issues worthy of further investigation. First, most of the results presented here are obtained by our numerics. But it is expected that at least some of them, like the two critical lines demonstrated in Fig.\ref{diagram} as well as the condensate near the critical lines, should be tractable analytically, albeit not readily. We are attempting to develop
the relevant analytical techniques to attack them. Second, as done for the bulk self-interacting scalar field in \cite{BG}, which appeared as our paper was being finalized, it is intriguing to scan the relevant parameter space in a meticulous manner to check whether the chaotic behavior also shows up in our periodically driven holographic superconductor since our bulk is also governed by the non-linear dynamics. Last but not least, the real time dynamics of  the conductivity, which is also of particular interest in the holographic superconductor, has yet to be investigated. Nevertheless as we know, unlike the condensate, which is an one-point correlation function, the conductivity is related to the two-point correlation function. The computation strategy for the two-point correlation function in the dynamical setting has been well developed in \cite{Balasubramanian3}, although to work it out explicitly turns out to be a non-trivial task. We hope to report our work on these issues in the near future.

\section*{Acknowledgements}
We would like to thank Zhoujian Cao, Ben Craps, Michael Crossley, Sijie Gao, Yi Ling, Jinshan Wu, Xiaoning Wu, and Xiaoxiong Zeng for their enjoyable discussions related to AdS/CFT correspondence in the dynamical setting. We also benefit greatly from the correspondence with Sean Hartnoll and Matthew Lippert on the determinant method in identifying quasi-normal modes. Brian Higgins, Xiaomei Kuang, and Chao Niu are acknowledged for their communications relevant to our numerics. Nice comments and suggestions from Julian Sonner are also much appreciated. W.J.L is grateful to Hong Liu for his encouraging discussions with various valuable suggestions. His work is financially supported by the China Scholarship Council. He is also partially supported by U.S. Department of Energy(D.O.E.) under cooperative research agreement DE-FG0205ER41360, NCET-12-0054, and NSFC Grants No.10975016, 11235003 as well as 11275208. Y.T. is supported by NSFC Grant No.11075206.
H.Z. is supported in part
by the Belgian Federal Science Policy Office through the
Interuniversity Attraction Pole P7/37, by FWO-Vlaanderen through
the project G.0114.10N, and by the Vrije Universiteit Brussel
through the Strategic Research Program "High-Energy Physics". He would also like to thank the financial support from the European Science Foundation through the HoloGrav Network and the wonderful hospitality from APCTP for his attending the Focus Program on holography, where the revision of this work is made.

\section*{Appendix: Gauge invariant formulation of bulk dynamics}
This appendix is devoted to the reformulation of bulk dynamics in terms of gauge invariant quantities $\Phi=|\Psi|$ and $M=A-\nabla \mathrm{arg}(\Psi)$, which turns out to be somewhat more convenient when one is dealing with the linear perturbation theory of bulk dynamics associated with the bulk solution dual to the boundary superconducting phase. As such, by rewriting the bulk action in favor of the above gauge invariant quantities and making a variation, we end up with the following equations of motion
\begin{equation}
 \mathcal{D}_a\mathcal{D}^a\Phi+(\mathcal{D}_a\mathcal{D}^a)^*\Phi-2m^2\Phi=0, \nabla_aF^{ab}=i(\Phi \mathcal{D}^b\Phi-\Phi \mathcal{D}^{b*}\Phi)=2M^b\Phi^2,
 \end{equation}
 where $\mathcal{D}=\nabla-iM$. Furthermore,  with $\Phi=z\chi$, the equations of motions can be explicitly written as follows
\begin{eqnarray}
0&=&2\partial_t\partial_z\chi+\frac{2}{z^2}f\chi-\frac{f'}{z}\chi-f'\partial_z\chi-f\partial_z^2\chi-\partial_x^2\chi-\partial_y^2\chi\nonumber\\
&&+M_x^2\chi+M_y^2\chi-2M_tM_z\chi+fM_z^2\chi-\frac{2}{z^2}\chi
\end{eqnarray}
for the Klein-Gordon equation and
\begin{equation}
\partial_z^2M_t-\partial_z\partial_xM_x-\partial_z\partial_yM_y+\partial_x^2M_z+\partial_y^2M_z-\partial_t\partial_zM_z=2M_z\chi^2,
\end{equation}
\begin{eqnarray}
&&\partial_t\partial_zM_t+\partial_t\partial_xM_x+\partial_t\partial_yM_y-\partial_x^2M_t-\partial_y^2M_t-f\partial_z\partial_xM_x-f\partial_z\partial_yM_y\nonumber\\
&&+f\partial_x^2M_z+f\partial_y^2M_z-\partial_t^2M_z=-2M_t\chi^2+2fM_z\chi^2,
\end{eqnarray}
\begin{eqnarray}
&&\partial_z\partial_xM_t+f\partial_z^2M_x+f'\partial_zM_x-2\partial_t\partial_zM_x+\partial_y^2M_x-\partial_x\partial_yM_y\nonumber\\
&&-f'\partial_xM_z-f\partial_x\partial_zM_z+\partial_t\partial_xM_z=2M_x\chi^2 \  \ (x\leftrightarrow y)
\end{eqnarray}
for the Maxwell equations. With the ansatz we have made for our purpose, i.e.,
\begin{equation}
M_y=\partial_xM_t=\partial_yM_t=\partial_xM_z=\partial_yM_z=\partial_xM_x=\partial_yM_x=\partial_x\chi=\partial_y\chi=0,
\end{equation}
the equations of motion can be reduced to
\begin{eqnarray}
&&2\partial_t\partial_z\chi+\frac{2}{z^2}f\chi-\frac{f'}{z}\chi-f'\partial_z\chi-f\partial_z^2\chi+M_x^2\chi-2M_tM_z\chi+fM_z^2\chi-\frac{2}{z^2}\chi=0,\\
&&\partial_z^2M_t-\partial_t\partial_zM_z=2M_z\chi^2,\\
&&\partial_t\partial_zM_t-\partial_t^2M_z=-2M_t\chi^2+2fM_z\chi^2,\\
&&f\partial_z^2M_x+f'\partial_zM_x-2\partial_t\partial_zM_x=2M_x\chi^2, 
\end{eqnarray}
 whereby the equations for the linear perturbation are given by
 \begin{eqnarray}
&&2\partial_t\partial_z\delta\chi+\frac{2}{z^2}f\delta\chi-\frac{f'}{z}\delta\chi-f'\partial_z\delta\chi-f\partial_z^2\delta\chi+2M_x\chi\delta M_x+M_x^2\delta\chi\nonumber\\
&&-2M_z\chi\delta M_t-2M_t\chi\delta M_z-2M_tM_z\delta\chi+2fM_z\chi\delta M_z+fM_z^2\delta\chi-\frac{2}{z^2}\delta\chi=0,\\
&&\partial_z^2\delta M_t-\partial_t\partial_z\delta M_z=2\chi^2\delta M_z+4M_z\chi\delta\chi,\\
&&\partial_t\partial_z\delta M_t-\partial_t^2\delta M_z=-2\chi^2\delta M_t-4M_t\chi\delta\chi+2f\chi^2\delta M_z+4fM_z\chi\delta\chi,\\
&&f\partial_z^2\delta M_x+f'\partial_z\delta M_x-2\partial_t\partial_z\delta M_x=2\chi^2\delta M_x+4M_x\chi\delta\chi.
\end{eqnarray} 
In the time averaged approximation, the final state can be approximated by
\begin{equation}
\chi=\chi^f(z),M_t=M_t^f(z),M_z=M_z^f(z)=\frac{M_t^f}{f},M_x=-\frac{E}{\omega}\sin(\omega t),
\end{equation}
where $\chi^f$ and $M_t^f$ satisfy
\begin{eqnarray}
&&\frac{2}{z^2}f\chi^f-\frac{f'}{z}\chi^f-f'\partial_z\chi^f-f\partial_z^2\chi^f+\frac{E^2}{2\omega^2}\chi^f-\frac{M_t^{f2}}{f}\chi^f-\frac{2}{z^2}\chi^f=0,\\
&&\partial_z^2M_t^f=2\frac{M_t^f}{f}\chi^{f2},
\end{eqnarray}
with the boundary conditions
\begin{equation}
\chi^f=0, M_t^{f'}=-\rho
\end{equation}
at the AdS boundary and
\begin{equation}
M_t^f=0, (1+\frac{E^2}{2\omega^2})\chi^f+3\chi^{f'}=0
\end{equation}
on the horizon. On top of such a final state, the corresponding equations for our quasi-normal modes are given by
\begin{eqnarray}
&&-2i\Omega\partial_z\delta\chi+\frac{2}{z^2}f\delta\chi-\frac{f'}{z}\delta\chi-f'\partial_z\delta\chi-f\partial_z^2\delta\chi+\frac{E^2}{2\omega^2}\delta\chi\nonumber\\
&&-2M_z^f\chi^f\delta M_t-fM_z^{f2}\delta\chi-\frac{2}{z^2}\delta\chi=0,\\
&&\partial_z^2\delta M_t+i\Omega\partial_z\delta M_z=2\chi^{f2}\delta M_z+4M_z^f\chi^f\delta\chi,\\
&&-i\Omega\partial_z\delta M_t+\Omega^2\delta M_z=-2\chi^{f2}\delta M_t+2f\chi^{f2}\delta M_z,\\
&&f\partial_z^2\delta M_x+f'\partial_z\delta M_x+2i\Omega\partial_z\delta M_x=2\chi^{f2}\delta M_x.
\end{eqnarray} 
Note that $\delta\chi$ is generically coupled with $\delta M_t$ and $\delta M_z$ albeit decoupled from $\delta M_x$. So one is required to solve the above coupled equations for the quasi-normal modes associated with the scalar part by imposing $\delta\chi=0$ at the AdS boundary.


\begin{thebibliography}{99}
\bibitem{CY}P. M. Chesler, and L. G. Yaffe, Phys. Rev. Lett. 102,  211601(2009)[arXiv:0812.2053].

\bibitem{MKT}K. Murata, S. Kinoshita, and N. Tanahashi, JHEP 1007, 050(2010)[arXiv:1005.0633].

\bibitem{DNT}S. R. Das, T. Nishioka and T. Takayanagi, JHEP 1007,  071(2010)[arXiv:1005.3348].

\bibitem{AJ}T. Albash and C. V. Johnson,  New J. Phys. 13, 045017(2011)[arXiv:1008.3027].

\bibitem{EH}H. Ebrahim and M. Headrick,  arXiv:1010.5443.

\bibitem{Balasubramanian1}V. Balasubramanian, {\it et al.},  Phys. Rev. Lett. 106, 191601(2011)[arXiv:1012.4753]. 

\bibitem{Balasubramanian2}V. Balasubramanian, {\it et al.},  Phys. Rev. D84, 026010(2011)[arXiv:1103.2683].

\bibitem{HJW1}M. P. Heller, R. A. Janik, and P. Witaszczyk, Phys. Rev. Lett. 108, 201602(2012)[arXiv:1103.3452].

\bibitem{GPZ}D. Garfinkle and L. A. Pando Zayas, Phys. Rev. D84, 066006(2011)[arXiv:1106.2339].

\bibitem{BD} P. Basu and S. R. Das,  JHEP 1201, 103(2012)[arXiv:1109.3909].

\bibitem{BBCCF}V. Balasubramanian, A. Bernamonti, N. Copland, B. Craps and F. Galli, Phys. Rev. D84, 105017(2011)[arXiv:1110.0488].

\bibitem{KKVT}V. Keranen, E. Keski-Vakkuri and L. Thorlacius,  Phys. Rev. D85, 026005(2012)[arXiv:1110.5035].

\bibitem{HPG}H. Bantilan, F. Pretorius, and S. S. Gubser, Phys. Rev. D85, 084038(2012)[arXiv:1201.2132].

\bibitem{HJW2}M. P. Heller, R. A. Janik, and P. Witaszczyk, Phys. Rev. D85, 126002(2012)[arXiv:1203.0755].

\bibitem{GS}D. Galante and M. Schvellinger, JHEP 1207, 096(2012)[arXiv:1205.1548].

\bibitem{CK} E. Caceres and A. Kundu, JHEP 1209, 055(2012)[arXiv:1205.2354].
 
\bibitem{BLM} A. Buchel, L. Lehner and R. C. Myers,  JHEP 1208, 049(2012)[arXiv:1206.6785].

\bibitem{BGSSW} M. J. Bhaseen, J. P. Gauntlett, B. D. Simons, J. Sonner and T. Wiseman, Phys. Rev. Lett. 110, 015301(2013)[arXiv:1207.4194]. 

\bibitem{BDDN}P. Basu, D. Das, S. R. Das and T. Nishioka,  arXiv:1211.7076. 

\bibitem{ACL}A. Adams, P. M. Chesler, H. Liu, arXiv:1212.0281.

\bibitem{GGGZZ}X. Gao, A. M. Garcia-Garcia, H. B. Zeng and H. Q. Zhang, arXiv:1212.1049. 

\bibitem{BGS}W. H. Baron, D. Galante and M. Schvellinger, JHEP 1303, 070(2013)[arXiv:1212.5234].

\bibitem{CKY}E. Caceres, A. Kundu, D. L. Yang, arXiv:1212.5728.

\bibitem{Balasubramanian3}V. Balasubramanian, {\it et. al.}, arXiv:1212.6066.

\bibitem{BLMN}A. Buchel, L. Lehner, R. C. Myers and A. van Niekerk,  arXiv:1302.2924.

\bibitem{NNT}M. Nozaki, T. Numasawa, and T. Takayanagi, arXiv:1302.5703.

\bibitem{CKPT} E. Caceres, A. Kundu, J. F. Pedraza, and W. Tangarife, arXiv:1304.3398.

\bibitem{BDST}N. Bao, X. Dong, E. Silverstein, and G. Torroba, JHEP 1110, 123(2011)[arXiv:1104.4098].

\bibitem{HHH1}S. A. Hartnoll, C. P. Herzog, and G. T. Horowitz, Phys. Rev. Lett. 101, 031601(2008)[arXiv:0803.3295].

\bibitem{HHH2}S. A. Hartnoll, C. P. Herzog, and G. T. Horowitz, JHEP 0812, 015(2008)[arXiv:0810.1563].

\bibitem{AKL}I. Amado, M. Kaminski, and K. Landsteiner, JHEP 0905, 021(2009)[arXiv:0903.2209].

\bibitem{Leaver}E. W. Leaver, Phys. Rev. D41, 2986(1990).

\bibitem{FHS}D. Frederik, S. A. Hartnoll, and S. Sachdev, Phys. Rev. D80, 126016(2009)[arXiv:0908.1788].

\bibitem{PAA}D. Podolsky, A. Auerbach, and D. P. Arovas, Phys. Rev. B84, 174522(2011)[arXiv:1108.5207].

\bibitem{PS}D. Podolsky and S. Sachdev, Phys. Rev. B86, 054508(2012)[arXiv:1205.2700].

\bibitem{BG}P. Basu and A. Ghosh, arXiv:1304.6349.


\end{thebibliography}
\end{document}